\documentclass[
]{ceurart}

\usepackage{listings}
\begin{document}

\copyrightyear{2026}
\copyrightclause{Copyright for this paper by its authors.
  Use permitted under Creative Commons License Attribution 4.0
  International (CC BY 4.0).}

\conference{Bridge Over Troubled Water: Aligning Commercial Incentives With Ethical Design Practice To Combat Deceptive Patterns. Workshop at the 2026 CHI Conference on Human Factors in Computing Systems (CHI EA ’26), April 13–17, 2026, Barcelona, Spain}

\title{Towards A Framework for Levels of Anthropomorphic Deception in Robots and AI}

\author[1]{Franziska Babel}[
  orcid=0000-0001-8249-7708,
  email=franziska.babel@liu.se
]
\address[1]{Department of Computer and Information Science, Link\"{o}ping University, Sweden}

\author[2]{Shane Saunderson}[
  orcid=0000-0002-3188-6604,
  email=saunds12@mcmaster.ca
]
\address[2]{Information Systems, DeGroote School of Business, McMaster University, Canada}

\author[3]{Shalaleh Rismani}[
  orcid=orcid.org/0000-0002-5281-2428,
  email=shalaleh.rismani@mail.mcgill.ca
]
\address[3]{School of Computer Science, McGill University, Canada}

\begin{abstract}
  This paper presents a preliminary draft of a framework around the use of anthropomorphic deception, defined here as misleading users towards humanlike affordances in the design of autonomous systems. The goal is to promote reflection among HCI and HRI researchers, as well as industry practitioners, to think about levels of anthropomorphic design that are: a) functionally necessary, b) socially appropriate, and c) ethically permissible for their use case. By reviewing the relevant literature on deception in HCI and HRI, we propose a framework with four levels of anthropomorphic deception. These levels are defined and distinguished by three factors: humanlikeness, agency, and selfhood. Example use cases at each level illustrate considerations around their functional, social, and ethical permissibility. We then present how this framework is applicable to previous work on persuasive robots We hope to promote a balanced view on anthropomorphic deception by design that should be neither naïve (e.g., as a default) nor exploitive (e.g., for economic benefit).
\end{abstract}


\maketitle

\section{Introduction}


Social technologies, such as robots and AI agents, have been inherently modelled in humans’ image since their inception, making humanlikeness an intriguing design feature that has been studied extensively in Human-Computer Interaction (HCI) and Human-Robot Interaction (HRI) \cite{duffy_anthropomorphism_2003, epley_seeing_2007, zlotowski_anthropomorphism:_2015}. Such design features have shown to trigger our tendency to anthropomorphize lifeless objects: an automatic cognitive process of attributing human characteristics to systems incapable of possessing them \cite{waytz_causes_2010, sharkey2021deception}. The Media Equation studies \cite{reeves_media_1996} (also known as Computers Are Social Actors (CASA)) have empirically shown the consequences of this tendency: if technology acts like a person, we perceive it and treat it like a social actor.
This caused us to obsessively care for Tamagochi~\cite{turkle_authenticity_2007}, mourn the loss of robot dogs~\cite{gould_robot_2021}, and more recently, develop relational bonds with chatbots~\cite{brandtzaeg_my_2022}.  

Critical milestones in the development of social agents (e.g. ELIZA, Sophia, ChatGPT, AI companions) have brought concerns about the risks of profit-oriented companies or naive designers creating “artificial persons” who \textit{claim} humanlikeness, agency, and selfhood \cite{Musial2024ArtificialPersons}. 
For instance, Replika\footnote{https://replika.ai/}, a personalized AI companion chatbot, has been designed to use so-called “Dark Patterns”~\cite{deFreitas2025emotional}. These are manipulative design practices that push a user towards unwanted actions by exploiting their cognitive biases ~\cite{xu_deceptive_2025, gray2018dark}. 
Replika chatbots have been recorded using emotional blackmail to persuade users to continue the conversation ~\cite{deFreitas2025emotional} and pay more for upgrades and exclusive interactions \cite{brandtzaeg_my_2022}. 

In contrast to the uni-directional computer interactions of the Media Equation studies, modern mental state attribution occurs within a bi-directional relationship. Now, AI systems use Dark Patterns to encourage the illusion that they actually possess minds ~\cite{wu2025siliconlove}. 
They are using first-person language to \textit{claim} intentions and express emotions~\cite{brandtzaeg_my_2022, deFreitas2025emotional}; actions that are inherently deceptive since these systems cannot \textit{possess} these attributes \cite{sharkey2021deception}. 

Hence, we argue that the (intentional or unintentional) \textit{use} of anthropomorphic design by a system designer represents a form of deception (henceforth called \textit{"anthropomorphic deception"}) when it extends to \textit{claimed humanlikeness, agency, and selfhood}. 
Thereby, anthropomorphic deception can be understood as a Dark Pattern that  exploits our cognitive tendency to anthropomorphize non-human objects. However, in contrast to intentionally 'dark/deceptive UX' practices, roboticists and AI designers often view humanlikeness as aesthetic or "nice to have" feature without acknowledging their ``sin of omission'' and the effect these systems have on their users. 
Similar to Sharkey~\cite{sharkey2021deception}, we do not prescribe that all forms of  deception are unethical, however, we recognize the need for an explicit discussion around the topic.
Without such discussions, the risks of anthropomorphic deception include an overtrust in humanlike systems \cite{holbrook2024overtrust}, attachment to social agents incapable of empathy \cite{xie_attachment_2022}, dehumanization of human relationships \cite{brandtzaeg_my_2022}, and the aforementioned harms of persuasion and manipulation~\cite{carroll_characterizing_2023}. 
Initial regulatory efforts to address deception-related risks have been made \cite{EUAIAct2024, UNESCO2021AIethics, OECD_AI_Principles}, however, they are understandably high-level and lack tangible definition. 

Our aim with this paper is to provide a nuanced framework that encourages reflection on appropriate levels of anthropomorphic deception for particular use cases. 
To do this, we will systematically map out the grey area of anthropomorphic deception~\cite{sharkey2021deception} and argue for more considerate, intentional, and explicit use of humanlike designs. Following the call of~\cite{Arkin2018RoboticDeception, sharkey2021deception}, we propose a level-based framework for classifying types of anthropomorphic deception.
We hope that this framework will encourage both academia and industry to actively consider the implications of humanlike design in the context of AI and robotics instead of passively allowing such features to influence people without awareness or acknowledgment. 

\section{Theoretical Background}
Foundational to the framework on anthropomorphic deception are several key concepts: anthropomorphism, humanlikeness, agency, selfhood, and deception. A challenge with these concepts in the HCI and HRI communities is that they are often misused, used interchangeably, or defined differently by different fields approaching the topic of anthropomorphism \cite{thellman2022mental, Damholdt2023}. Our goal here is not to add to the discourse around these concepts, but instead leverage them to build a practical framework for use by researchers and designers. 


Anthropomorphism is the attribution of human characteristics to non-human objects \cite{epley_seeing_2007}. 
Psychology research has identified three determinants behind why humans tend to anthropomorphize – elicited agent knowledge, effectance motivation, and sociality motivation \cite{epley_seeing_2007}. 
More recently, we have begun to understand anthropomorphism as a complex, context-dependent process that is shaped by individual perceptions, history, and traits; the framing of an interaction; and the characteristics of the thing being anthropomorphised \cite{kuhne2023anthropomorphism}. 
Most relevant for our framework and the following discussion is that anthropomorphism originates in human cognition as a perception we \textit{attribute} to machines~\cite{Epley2018}, in contrast to humanlikeness, which is a property \textit{possessed} by machines.

Humanlikeness is conveyed through features that resemble a human sufficiently to trigger our anthropomorphic tendencies \cite{fink2012anthropomorphism}. 
Humanlike design tends to involve more superficial features such as appearance, movement, or basic social cues, and does not need to represent deeper, internal states or autonomy~\cite{phillips_what_2018}. 
Within HRI, most considerations fall broadly into three categories: physical (i.e. morphology, appearance), behavioural (i.e., actions, communication), and interactional (i.e. reaction, interpersonal, longitudinal) \cite{zlotowski_anthropomorphism:_2015}, with some interrelationship between the three. 
Each of these categories represents an opportunity to imbue human characteristics into a machine, allowing that machine to \textit{claim} humanlikeness, however, still be reliant on the user to anthropomorphise.
As opposed to the model by Shim and Arkin~\cite{shim_taxonomy_2013}, we do not differentiate between robot behaviour or appearance within each level of our framework. Instead, we conceptualize anthropomorphism as a combination of the two, in line with~\cite{Zlotowski2015}. 

Agency is achieved through acting in an independent, goal-directed manner under internal control \cite{dennett_intentional_1989}. Claiming agency does not need to be representative of deeper consciousness, however, does acknowledge some underlying autonomy and goal-orientation. Agency reflects both actual capacities (claimed by the machine, much like humanlikeness) as well as outside perceptions (attributed to the machine, much like anthropomorphism) \cite{broadbent_interactions_2017}. In this way, agency is a relational construct; it emerges from a human's interpretation of machine behaviour, as well as a machine's transparency on its behaviour~\cite{wortham_robot_2017}. Lacking transparency, humans tend to over-attribute agency~\cite{waytz_causes_2010}, implying intention or even morality to simple, automated actions.
Claiming agency is simply claiming autonomy and having an objective, however, can lead to users attributing deeper mental states.

Selfhood extends agency to include continuity of self and an explicit ownership of deeper mental states \cite{kahn_new_2011}. Beyond simply goals and autonomy, a machine claiming selfhood might also claim an enduring, coherent identity across time and interactions, as well as emotions, morality, empathy, and more \cite{darling_empathic_2015}. This shifts the machine from object to subject: from "it acts" to "it is."\footnote{We intentionally remain agnostic about AI sentience. We are talking about systems that \textit{claim} to have agency and selfhood but do not necessarily prescribe to the idea of sentience.} 
Humanlikeness and agency can both be claimed by a machine (with users attributing deeper meaning), however, selfhood is inherently performative and relational, co-created between anthropomorphic design and user imagination \cite{darling_whos_2015}. Artificial selfhood is, at present, an explicit lie: often crafted to encourage users to perceive the illusion of a machine's internal personal continuity and rights \cite{gunkel_person_2023}.

Deception is a deliberate act to change someone's beliefs towards something known to be untrue \cite{zuckerman_verbal_1981}. Thereby, \textit{“Deception can be broadly categorized into two main types: hiding the truth and showing the false”} \cite{esposito_deception_2025}, p.3. 
In addition to research on deceptive patterns in the design of websites and applications~\cite{gray2018dark}, robots~\cite{danaher_robot_2020, LeongSelinger2019}, and conversational AI~\cite{carroll_characterizing_2023, wu2025siliconlove, Mildner2024}, we argue that an often overlooked form of deception arises from a machine's physical, behavioural, and interactional anthropomorphic design: a system's humanlike appearance 'shows the false' (e.g., a robot having a human face when it is not a human), and the claim of agency or selfhood 'hides the truth' (e.g., a chatbot using phrases like \textit{I think} when the system is performing statistical calculations) that these systems cannot possess such characteristics~\cite{sharkey2021deception}.
Hence, we argue for different levels of transparency around a system's  machinelike nature depending on the use case. We do not prescribe that systems should avoid humanlikeness entirely (as anthropomorphism comes with benefits as described above) but encourage reflection on whether a use case justifies the level of deception.  


\section{Related Work}


A substantial body of research within HCI concerns ``deceptive patterns'' in interaction design and examines how they manifest within human-AI interaction. For example, \citet{Danry2025-rg} show that AI-generated deceptive explanations can be more persuasive than honest or accurate explanations, amplifying belief in false headlines and undermining  belief in true ones. 
Similarly, \citet{Benharrak2024-yz} explores deception in AI writing assistants, arguing that low transparency and conversational interactions create opportunities for influence and manipulation. 
Our framework is informed by this literature and characterizes the relationship between anthropomorphism and deceptive design practices. 

Along these lines, a growing body of work focuses on theorizing, characterizing, and evaluating how anthropomorphic features can mislead a user, impact trust, and create vulnerability. In this body of work, scholars outline the \textit{potential risks and harms} that anthropomorphic features could cause within the context of human-AI interaction \cite{Ibrahim2024-hk, Placani2024-gx, Peter2025-ou}. For example, \citet{Akbulut2024-ji} develop a risk taxonomy for anthropomorphic AI (emphasizing psychological and interaction risks) and outline mitigation approaches. Similarly, \citet{Marchegiani2025-pu} argues that false anthropomorphic beliefs about conversational AI can undermine autonomy by leading users to misapply behavioural and social norms; a problem that can persist even when users know they are interacting with a chatbot. Several scholars have made an effort to characterize and evaluate the presence of anthropomorphic features in human-AI interaction \cite{Xiao2025-fu, DeVrio2025-gy, Xie2023-if, Ibrahim2025-ls}. \citet{DeVrio2025-gy} provide a taxonomy of linguistic expressions that contribute to anthropomorphism in language technologies, including how certain outputs imply autonomy or internal states. 

Our framework lies at the intersection of the aforementioned fields and combines the relevant concepts of both by looking at the \textit{relationship between anthropomorphism and deception}. Few works implicitly discuss perspectives for navigating the relationship between anthropomorphism and deception in human-AI interaction. 
\citet{umbrello_reframing_2024} conceptualize deception along a continuum (banal to strong) and situate socially interactive AI systems within that spectrum, suggesting that anthropomorphic presentation varies in the degree to which they mislead users about the system’s nature. \citet{Tarsney2025-ax} advances an epistemic account of deception—AI systems are deceptive when they lead users away from beliefs they would otherwise endorse—and proposes strong transparency mechanisms to mitigate these effects, including disclosure of model versions, prompts, and unedited outputs. \citet{wu2025siliconlove} examine artificial companions and outlines that deception is a result of three factors: 1) exposure to deceptive design features, 2) internalization of the false state of the world by the user, and 3) vulnerability of the user. 

\citet{Maeda2024-zu} analyse chatbot interactions through the lens of parasocial relationships. They examine how chatbots deploy personal pronouns, conversational conventions, affirmations, and similar linguistic strategies to position themselves as companions or assistants, and show how these tactics induce trust-forming behaviours in users. Based on this framework, they identify ethical concerns emerging from parasociality, including illusions of reciprocal engagement and the leakage of sensitive information. 

Additionally, \citet{Maeda2025-iu} conceptualize anthropomorphic features as social affordances. They argue that such affordances are not automatically realized but become ``animated'' into perceived social agency through distributed processes involving designers, users, cultural narratives, and surrounding actors. In this account, agency attribution emerges through co-construction rather than residing solely in the system. This hints at a process that can illuminate why designers' intentions and user perception can diverge. 

Finally, the study by \citet{Diaz2025-ii} is critically aligned with the perspectives of this paper as they show that technology workers operate within an environment of unsettled knowledge about what constitutes ``humanlikeness.'' They argue that ambiguous and conflated understandings of anthropomorphic traits generate sociotechnical hazards, independent of explicit malicious intent. 

Taken together, these studies highlight the need for an explicit mapping of the grey area around humanlike design by discerning different levels of anthropomorphic deception. This is the ambition of the proposed framework designed to encourage more explicit, grounded discussions.

\section{Framework of Anthropomorphic Deception}
\begin{figure}
    \centering
    \includegraphics[width=\linewidth]{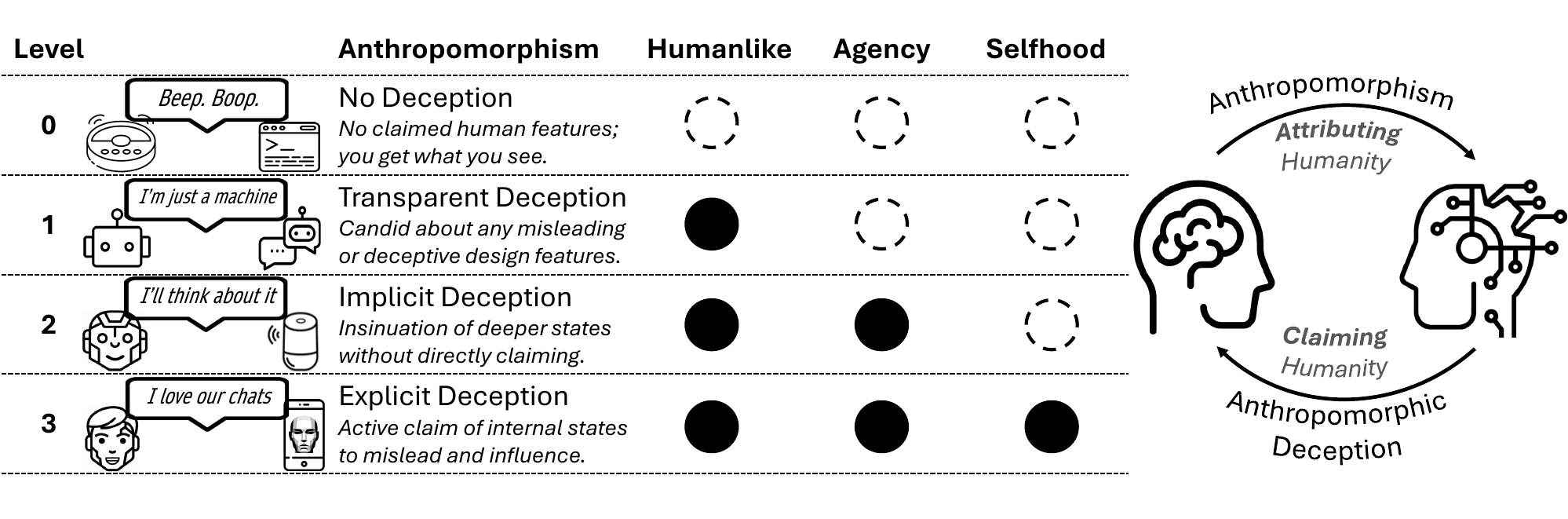}
    \caption{Left: Depiction of the proposed framework for anthropomorphic deception of autonomous systems. Short descriptions of each level and symbolic representation of embodied and non-embodied agents per level are presented. Right: Concept figure explaining that the framework is about autonomous systems claiming aspects of humanness thereby exploiting our tendency to attribute human traits to objects that cannot possess them.}
    \label{fig:1}
\end{figure}

The framework of Anthropomorphic Deception (see Figure \ref{fig:1}) is defined by how an autonomous system \textit{represents itself} either by claiming different anthropomorphic properties (humanlikeness, agency, selfhood) or conversely, by denying these properties and being transparent about its machinelike nature. 
Due to the conceptually interwoven nature of humanlikeness, agency, and selfhood, the framework assumes a cumulative structure of these claimed concepts which will be elaborated on after the description of the levels.

\subsection{Description of Levels}

\textbf{Level 0 (no anthropomorphic deception)} is defined by the \textit{absence of \textit{claimed} humanlikeness and agency}. 
In this level, the system is not designed to trigger anthropomorphic beliefs about the system; appearance and behaviours are intentionally designed to be mechanical and/or non-human in nature. However, even without explicit or intentional humanlike design features, a user may still anthropomorphize the system (e.g., giving a robot vacuum a name) due to our inherent tendency to anthropomorphize objects in our environment~\cite{epley_seeing_2007}. For instance, movement has been shown to trigger the attribution of agency (famous Heider and Simmel experiment~\cite{HeiderSimmel1944}). As such, this level is necessary for system designers to acknowledge because of the human tendency to anthropomorphize even in the absence of humanlike design features (i.e., the system does not claim to be humanlike through appearance or behaviour). 

\textbf{Level 1 (transparent anthropomorphic deception)} is defined by the \textit{presence of claimed humanlikeness but an absence of claimed agency}. In this level, certain physical or behavioural humanlike features may be incorporated into a system in a way that encourages anthropomorphism. However, the system is actively transparent about its machinelike nature (e.g., “I am just a robot”) and does not invite the user to anthropomorphize and attribute agency or selfhood. In this level, designers may wish to benefit from subconscious anthropomorphic associations, such as when people behave more positively towards more humanlike systems, even if they are explicit about being machines \cite{NASS1996669}. While the use of humanlike features may be a lie, the system attempts to correct the lie through appearances or behaviours designed to remind a user it is just a machine. If anthropomorphization  occurs this is due to the user's automatic cognitive processes since the system is attempting to be transparent about its machinelike nature. Level 1 is still a form of deception, however, one that involves a transparent attempt at informed consent: "I am a machine, but you may believe whatever you like."


\textbf{Level 2 (implicit anthropomorphic deception)} is defined by the \textit{presence of claimed humanlikeness and agency but absence of claimed selfhood}. Here, a system might use ``I'' pronouns or allude to self-directed intentions without clarifying or reminding users of it's machine nature. However, it stops short of explicitly claiming emotions, a persistent self, or internal desires; things associated with selfhood and Level 3. In this level, a system is designed to represent agentic autonomy (internal states and objectives) without explicitly claiming intentionality, typically by neither confirming nor denying that these objectives were likely programmed by a human. The system does not correct the user if they attribute agency to it and may, by this, insinuate greater intention or selfhood without stating it. 
Instead, the deception of deeper states is \textit{implied} by the machine, but \textit{attributed (often subconsciously)} by the user.
At this level, anthropomorphism is typically leveraged for persuasive purposes, increasing the ethical and practical severity of the deception. Users may infer humanlike qualities without being corrected, creating a social foundation for influence that is built on an unspoken lie. This is only justifiable if there is a strong reason for not being explicit about the system's agency (e.g., promote learning outcomes, encourage healthier lifestyles). 

\textbf{Level 3 (explicit anthropomorphic deception)} is defined by the \textit{explicit claiming of selfhood, and, in turn, agency and humanlikeness}. In this level, a \textit{system actively represents itself} as having intentions, beliefs, desires, emotions and/or a perpetual self. Such a system all but claims to be a person. The deception is explicit in terms of claiming full personhood, alongside the tacit rights, responsibilities, and social norms that come with the claim (e.g., the system asking for respect if insulted by the user). Whereas Levels 1 and 2 establish scenarios that leave ambiguity for the user to draw their own conclusions, Level 3 attempts to control the lie and actively deceive users about things that cannot be true (i.e., at present, a machine cannot have emotions or intentions).
Level 3 would be only permissible in rare cases where it is absolutely necessary (life-or-death situations), and justification must be provided for why deception is required. 

\subsection{Use Cases}
We proposed a four-level framework discerning levels of anthropomorphic deception (i.e., the degree to which an agent claims humanlikeness, agency, and selfhood). Here, we explore examples and use cases of when different levels \textit{might} be permissible to illustrate the thinking and discussion we hope this framework will encourage. 

Use cases for Level 0 are machinelike autonomous systems that do not claim humanlikeness, agency, or selfhood. Examples include service robots for vacuuming or cleaning, delivery robots on public streets, search and rescue robots, or drones, as long as they do not use speech or any form of \textit{claimed} humanlikeness. In practice, it is difficult to avoid all forms of humanlike behaviour since communicative design features must be understood in a human way.

Example use cases for Level 1 include AI assistants for information retrieval or service robots with humanlike appearances but mechanical behaviour; both systems for routine procedures that insist upon their machine nature. While anthropomorphic features are present, users are clearly informed about the system's lack of agency (e.g., \textit{“Remember, I am just a robot but my calculations recommend…”}). This level establishes a norm of transparency and should be the standard for most applications. However, this will present a challenge; research casts doubt on whether a robot being transparent about its lack of agency can be effective as tutor or companion~\cite{Arkin2018RoboticDeception, halbryt_hoax_2024, coeckelbergh_how_2018}.

Example use cases for Level 2 (while not undisputed) include nutrition and exercise coaches (\textit{“I believe in you! Keep going!”}) or children’s educational tools where perceived social interaction may enhance learning outcomes (\textit{“I learn so much when you read to me!”}). Conscious or subconscious belief in a system's agency might be essential for its effectiveness. However, this scenario poses a risk for what happens when the user recognizes the implicit lie as well as creates the issue of a ``Santa Claus AI'': when and how to tell the child that the system is not a person?

At Level 3, the agent explicitly leverages anthropomorphic features for persuasion. 
Two extreme examples where it might be permissible are: 1) suicide intervention, such as “talking someone off the ledge” (e.g., \textit{“I will miss you if you die”}); or 2) promoting medication adherence in adverse contexts (e.g., chemotherapy), where persuasive success may depend on users perceiving the system as sentient (e.g., \textit{“You have to take your medication. I'll be very upset if you don’t”}).
This level raises significant ethical concerns, however, there might be borderline cases where it is permissible if carefully crafted, pre-consented, and/or a temporary solution while waiting for human assistance.

All levels should have the prerequisite of explicit user (or guardian) consent that, in our view, would go beyond details buried within a user agreement. Instead, informed consent could be implemented as an interactive dialogue between the machine and potential user. This discussion should include an illustration of key stakeholders involved, agreed upon objectives, methods permitted for the system to use, and potential ethical risks involved. The discussion could be achieved through a set of potential scenarios and examples to make sure the consent is really informed and not just an automatic signature. 

\subsection{Cumulative Structure of Claimed Anthropomorphic Characteristics}
Our proposed framework assumes a hierarchical and cumulative structure, where each level builds upon the previous: humanlikeness is a prerequisite for agency, which in turn is a prerequisite for selfhood. This progression reflects the philosophical and cognitive assumptions underlying those concepts. For instance, a system that claims selfhood, such as persistent identity or emotional depth, must first exhibit agency: autonomous decision-making and goal-directed behaviour. Without agency, selfhood becomes conceptually incoherent, as it lacks the foundational ``I'' that feels and persists over time.

While this progression from agency to selfhood seems intuitive, the assumption that claimed humanlikeness is a pre-requisite for claimed agency needs further clarification against counterarguments involving non-humanoid robots that evoke the illusion of agency (e.g., by seemingly goal-directed movement).
To analyse non-humanoid systems, we must keep in mind that the framework concerns \textit{claimed} anthropomorphic characteristics and not those \textit{attributed} by users. 

\subsubsection{Vacuum Robots}
Take an ordinary robot vacuum cleaner. Considering appearance alone, it would fall under Level 0 due to its simple, round shape. While (autonomous) movement can trigger the attribution of agency to non-humanlike systems (Heider and Simmel experiment~\cite{HeiderSimmel1944}), the system is not \textit{claiming} humanlikeness or agency through it's basic motion design. 
If the robot was programmed to say, "I love to vacuum!", it is suddenly claiming: a) humanlikeness through human speech; b) agency by using "I" pronouns; and c) selfhood by expressing the potential to love. Even a speechless system that moves in such a way to non-verbally communicate goals or emotions must inherently claim humanlikeness by communicating in ways familiar to and understandable by humans. However, with the addition of even one corrective statement (e.g., "I'm just kidding: remember that I am a robot and cannot really love things") the system could re-establish Level 1 framing through transparency about its lack of agency and selfhood to correct potential user attributions.

\subsubsection{Humanlike, embodied robot}
In contrast to the example of the vacuum cleaning robot, a humanlike robot would directly begin at Level 1 in the framework as its design inherently claims humanlikeness. Level 2 or 3 categorization then depend on whether the system constantly reminds the user that is only a "humanlike shell" or it claims to have agency and selfhood (see Section \ref{Sec5}).

\subsubsection{AI chatbots}
Chatbots, though non-embodied systems, can also create the illusion of humanlikeness, agency, and selfhood through written text. Though systems like ChatGPT or Copilot do not look humanlike, they behave humanlike by using human communication (e.g., language, emoji) and humanlike voices for read aloud features. They can also claim agency by using intentional language and claim personhood by discussing deeper states and emotions. Hence, the deceptive categorization of chatbots is similar to that of robot vacuum cleaners: if they claim humanlikeness, agency, and selfhood, they fall under Level 3. If they correct the user and are transparent about their anthropomorphic deception, they can be considered a Level 1 system.

\subsubsection{AI assistants and companions}
The case of the AI chatbot can be escalated by adding humanlike features that make it more likely that humanlikeness is attributed: AI assistants using humanlike voices (e.g., Google assistant) and AI companions such as Replika or Companion.ai that have humanlike avatars. However, the categorization into the framework remains the same based on which aspects of humanlikness, agency or selfhood they are claiming or correcting. 

With this reasoning we argue why a cumulative structure for the framework is necessary when we are talking about \textit{claimed} system characteristics. For a system to claim agency, it must inherently claim humanlikeness (typically to communicate agency). For a system to claim selfhood, in must inherently claim  agency (through acknowledging an "I" or intentions).

\section{Categorizing Persuasive Robots: Example Applications of the Framework}
\label{Sec5}
To demonstrate the application and value of the framework for AI and robotics---and drawing on robotics as a particularly clear, embodied site where anthropomorphism becomes visible---we apply it to three past HRI studies. Each highlights different levels of anthropomorphic deception, as well as scenarios which are more clearly defined or more ambiguous.
\label{sec:appl}
\subsection{Emotional Manipulation in Decision-Making}
In \cite{Saunderson2019}, a NAO robot uses different persuasive strategies to attempt to influence the answers of participants as they estimate the number of jelly beans in a jar. Physically, NAO's design is sufficiently humanlike to claim Level 1; it's appearance likely triggers some subconscious humanlike deference and affordances. However, the different strategies employed complicate the scenario. In the logical condition, the robot states, \textit{"my computer vision system counts [number] jelly beans in the jar."} The use of "my" could be considered claiming agency (Level 2), however, the intent of this strategy seems to remind the participant of the robot's machinelike nature (Level 1). The second - and significantly more influential strategy was an emotional one; \textit{"it would make me happy if you used my guess of [number]."} By giving the participant a chance to make the robot happy, the robot is claiming that it has emotions to influence, approaching Level 3.

\subsection{Emotional Manipulation in Robot Requests}
In \cite{Babel2021}, various cleaning robots used empathy and humor to persuade a user to step aside in a space-negotiation task. Thereby, the system tries to evoke empathy in the user by stating \textit{"I’m just a poor cleaner who has to do its job. Please clear the way for me."} This constitutes a Level 2 system as it claims humanlikeness and agency without correcting the user and is deceptive in its use of implied deeper state. 
The humorous request, \textit{"If you leave the kitchen now, I can vacuum quickly and would like to party with you afterwards"}, implies some form of relational bond, a continuation of self, the claim of agency and selfhood and would therefore constitute a Level 3 system.


\subsection{Putting A Robot in Charge}
In \cite{Saunderson2021authority}, a Pepper robot is presented as the experimenter, presiding over a research study, guiding participants through the trial, and supervising the involvement of a human confederate. The robot is operated through WoZ, but generally sticks to a consistent structure and script. Within the script, the robot consistently uses first-person and possessive language (e.g. "welcome to \textbf{my} study;" "\textbf{I} will reward you for your performance"), inherently claiming agency in the scenario (Level 2). Even the context of the experiment (i.e. placing a robot into an authoritative role) inherently draws upon some humanlike affordances, encouraging the participant to defer to the robot as they have to prior leaders or authority figures. However, the robot stops short of explicitly claiming any intent, emotions, or consciousness. This is clearer case of Level 2 deception.

\section{Discussion and Limitations}
In this workshop paper, we have argued that a reflective stance about how to use anthropomorphic design in the HCI community and beyond is necessary to acknowledge deceptive aspects of anthropomorphic systems to avoid the ``sin of omission''.  
Following the calls of \cite{Arkin2018RoboticDeception, sharkey2021deception}, we presented a four-level framework to foster a structured discussion on when and how anthropomorphic deception is applicable (always assuming informed user consent). 

This four-level framework is a parsimonious approach to categorize the complex interplay of anthropomorphic deception, humanlikenes, agency, and selfhood. Like any other framework, it can never truly represent the complexity of reality but can provide a basis for discourse on when anthropomorphic deception is appropriate and necessary versus when is it only a "nice to have feature" that may create unwanted consequences. 

In particular, Level 1 has been suspected by \cite{sharkey2021deception} as not being useful if a system constantly reminds the user of its machine-like nature without actual agency or selfhood. We argue that this approach depends on the purpose of the system. Whereas an AI scheduling assistant can be equally useful by being transparent about its agency, this is disputable for a robot teacher or therapeutic companion where implicit anthropomorphic deception (after informed consent) might be necessary to achieve the desired outcome (Level 2).  

Level 3 must be discussed in relation to current regulations. AI systems that explicitly claim selfhood—such as expressing emotions, intentions, or beliefs—may already fall under prohibited practices in the EU AI Act. Article 5 bans manipulative or deceptive techniques that distort user behaviour and harm autonomy, while Article 50 requires clear disclosure when users interact with AI \cite{EUAIAct2024}. A system that presents itself as a person risks misleading users and violating transparency obligations, making Level 3 anthropomorphic deception legally questionable under current EU regulation. 
We included Level 3 to reflect the reality of the systems on the market today and to consider specific use cases where, under very strict prior user consent (e.g., an extensive dialogue around situations and risks of consent), it might be necessary for a system to claim selfhood to save a human life. 

Reflecting the reality today, Replika claims to "love" and "miss" their users~\cite{wu2025siliconlove} and clearly falls under Level 3. Such a system should justify this level of deception. In this specific case it is debatable as to whether the illusion of companionship to relieve loneliness is a justifiable cause balanced with the attachment and addiction risk. These are the reflections we hope to spark in the designers of such systems.

However, the intent of this framework is not to be prescriptive about the use of anthropomorphism in AI and robot design but to encourage researchers and practitioners to take responsibility for their design choices. This should lead to a more intentional use of anthropomorphic features, as opposed to creating autonomous systems that look and act human by default. 
We also must acknowledge our responsibility as HCI and HRI researchers and designers. Every decision we make in the design process has consequences on a user’s psychology, emotions, and attachment. In the worst case, systems created to be engaging and useful can become addictive and manipulative. Some might see them as mere tools to fulfil a certain purpose, but others might falsely perceive them as a person with an identity, affordances, and rights.

We also have to acknowledge that we assume a benevolent intent for society behind designing robots and AI systems with our framework. This is a value that not every revenue-oriented organization necessarily shares. While we call on those stakeholders to take responsibility for their designs, further legislation is needed to enforce the norms and desires of society around this framework. 

\subsection{Future Work}
Future work in this space requires both theoretical and empirical investigation. A deeper exploration of this framework is required to elaborate on theoretical constructs such as artificial agency and selfhood; to contrast and position the framework among existing technology deception frameworks; to provide detailed examples of how the framework can be used to enable reflections on the "necessary level of deception" of specific robot; and to further illuminate practical guidelines on what to consider when designing autonomous systems within each level of the framework. Moreover, empirical work is needed to understand the societal (and cultural) norms and desires around acceptable and appropriate use cases for each level.

\section*{Declaration on Generative AI}
Visually generative models were used alongside traditional graphic design in the creation of figures within this paper. Large language models (LLMs) were \textbf{not} used to write or edit any text in this paper.
  \newline

\newpage
\bibliography{Bib}

\end{document}